\documentclass[aps,prl,showpacs,twocolumn,psfig,epsfig]{revtex4}

\draft
\usepackage{hyperref}
\usepackage{graphicx}
\newcommand{\Om}{\Omega}

\newcommand{\gm}{\gamma}

\newcommand{\om}{\omega}

\newcommand{\ve}{\varepsilon}
\newcommand{\pa}{\partial}

\begin{document}

\title{Surface waves in uniaxially anisotropic left-handed materials}

\author{G. T. Adamashvili}
\altaffiliation[Permanent address: ]{Georgian Technical University, Kostava str.
77, Tbilisi, 0179, Georgia. }
\affiliation{Max-Planck-Institut f\"ur Physik Komplexer
Systeme, N\"othnitzer Str. 38, D-01187 Dresden, Germany}
\affiliation{Institut f\"ur Theoretische Physik, Nichtlineare
Optik und Quantenelektronik, Technische Universit\"at  Berlin,
Hardenbergstr. 36, D-10623 Berlin, Germany}
\author{A. Knorr}
\affiliation{Institut f\"ur Theoretische Physik, Nichtlineare
Optik und Quantenelektronik, Technische Universit\"at  Berlin,
Hardenbergstr. 36, D-10623 Berlin, Germany}

\begin{abstract}
The linear and nonlinear surface waves propagating along the interface separating isotropic conventional and uniaxially anisotropic left-handed materials is investigated. The conditions of the existence of  surface TM-modes is determined. It is shown that surface waves can be formed when the components of the permittivity and permeability tensors of the uniaxially anisotropic left-handed materials are simultaneously negative. A transition layer sandwiched between connected media is described using a model of a two-dimensional gas of quantum dots. Explicit analytical expressions for a surface optical soliton of self-induced transparency in the presence of single and biexciton transitions depends on the magnetic permeability of the left-handed medium, are given with realistic parameters which can be reached in current experiments. It is shown that  the sign of the total energy flow of the surface soliton depends on the parameters of the quantum dots and connected media.
\end{abstract}


%\Keywords{}

\maketitle

\centerline{I. Introduction}

The electromagnetic wave propagated  in isotropic regular media having simultaneously
positive  the electric permittivity $\ve$, and the magnetic permeability $\mu$, the vectors of the electric $\vec{E}$  and  magnetic $\vec{H}$ fields and wave vector $\vec{k}$ constitute a  right-handed triplet of vectors. In such media the
Poynting vector $\vec{S}$ is coincide with wave vector. Such media are sometimes labeled as  the ordinary right-handed materials (RM). When the plane electromagnetic wave propagation in an isotropic medium having simultaneously negative permittivity  $\ve<0$ and permeability $\mu<0$, should give rise to several other unusual electrodynamic properties compared with regular RM \cite{1}. While Poynting vector for
a plane wave still gives the direction of energy flow, the wave vector shall be in the opposite direction of $\vec{S}$ and vectors $\vec{E},\;\;$ $\vec{H}$ and $\vec{k}$ shall form a left-handed orthogonal set of vectors. In this case, they are termed left-handed materials (LM) or a negative refractive index materials (NIM). These artificial metamaterials are fascinating because have properties that do not occur in nature \cite{2,3,4}.

Pendry showed that a combination of split-ring resonators and metallic wires can lead to negative index of refraction \cite{5}. Starting with first realizations in the microwave region \cite{6}, recently, using silver as a constituent material, the achieved demonstrate a negative index of refraction at the red end of the visible spectrum \cite{7}.

The metamaterials that have been used in different experiments are usually anisotropic in nature, and very difficult to prepare an isotropic LM \cite{8}. The analysis of the symmetry of the LM used in Refs.\cite{9,10} it is shown that they have uniaxial anisotropy. The characteristic peculiarity of the electromagnetic waves propagating in the uniaxial anisotropic left-handed materials (ULM) is significantly  different than in isotropic LM. This distinction is especially evident in a layered structures when propagating waves pass from one isotropic right-handed materials (IRM) into another ULM and under these conditions of anomalous reflection or refraction are occur.  In addition, in general case in uniaxial anisotropic media, the vectors $\vec{E},\;\;$ $\vec{H}$ and $\vec{k}$ cannot form a strictly left-handed orthogonal set of vectors and the directions  of energy flow and the Poynting vector $\vec{S}$ cannot be in the exactly opposite directions of wave vector \cite{8}.

At the propagation of the electromagnetic waves along the interface in a layered structures the surface electromagnetic waves (SW) can be formed.  Surface modes can propagate in region of frequencies where the permittivities of two connected media have the opposite signs, i.e. SW propagate at the interface between media with different "rightness". Lately SW have attracted much interest in the context of nano-optics \cite{11}, diverse applications \cite{12} and also can be used as a convenient tool for studying properties of the structured surfaces which is considered as optical metamaterials \cite{13}. Characterized peculiarity of these waves are strong enhancement and spatial confinement of the
electromagnetic field of the wave near to the interface. In NIM interest to  SW  is connected with the strong impact of the SW on the image resolution of an LM flat lens \cite{14}. The properties of the non-resonance nonlinear SW propagating at the interface between conventional and LM  as well as between two different LM when medium to have an intensity-dependent dielectric permittivity in isotropic media are investigated \cite{15,16,17}.

Alongside with non-resonance nonlinear  waves exist as well another type the resonance nonlinear waves which can be formed within the McCall-Hahn mechanism, where a nonlinear coherent interaction takes place via Rabi-oscillations of the carrier density, if the conditions for self-induced transparency (SIT) are fulfilled:
\begin{equation}
\omega T>>1,\;\;\;\; T<<T_{1,2},
\end{equation}
in attenuator medium the steady-state $2\pi $ -pulse (soliton) is generated, when $\Theta >\pi$ \cite{18}. Here $T$ and $\omega $ are the width and the carrier frequency of the pulse, $T_{1}$ and $T_{2}$ are the longitudinal and transverse relaxation times of the atomic systems, $\Theta$  is the area of the pulse.

Besides the atomic systems in conventional media SIT for semiconductor quantum dots (SQD) also is realized. SQD, also referred to as zero-dimensional atoms (artificial atoms), are nanostructures which allow confinement of the charge carriers in all three spatial directions, resulting in atomic-like discrete energy spectra with strongly enhanced carrier lifetimes \cite{19}. Such features make quantum dots in many respects similar to atoms. SIT in SQD have been investigated experimentally \cite{20} and theoretically \cite{21,22} for plane waves and waveguide modes. SIT for SW in conventional media for atomic systems and for SQD have been investigated \cite{23,24,25}. In anisotropic media SIT has specific peculiarities which in conventional media for atomic systems have been studied \cite{26,27,28,29}.

At the propagation SW in anisotropic LM situation is absolutely different and have not considered up to now. The purpose of the present work is to theoretically investigate the processes of the formation of linear SW and surface optical solitons of  SIT   propagating along the interface between ULM and IRM.
At this for investigation of the surface solitons we consider the resonance transition layer of two-dimensional gas of inhomogeneously broadened SQD sandwiched on the ULM/IRM interface.

\vskip+0.5cm

\centerline{II.Basic equations}

We consider the propagation of the surface TM-modes along interface between the conventional IRM and the ULM in the case when optical pulse with width $T$ and frequency $\omega>>T^{-1}$ is propagating along positive direction of the $z$ axis. On the flat border of the division (x=0) between two connected media, a thin transition layer with thickness $d$ containing a small concentration of SQD having the polarization
$$
\vec P(x,z,t)=\vec e_{p}\;p(z,t)\;\delta(x),
$$
where $\vec e_{p}$ is the polarization unit vector along $z$ axis. Semi-spaces are divided at $x<0$ and $x>0$, with IRM (medium I) with electric permittivity $\ve_{1}(\om)>0$ and the magnetic permeability  $\mu_{1}(\om)>0$ and ULM (medium II), respectively. If we take the optical axis of the ULM $O$  parallel to the interface of two media along the $z$ axis and we assume that both the electric permittivity $\hat{\ve}$ and the magnetic permeability  $\hat{\mu}$ tensors are uniaxially anisotropic than we have \cite{30,31}:
$$
\hat{\ve}=\left(%
\begin{array}{ccc}
   \ve_{xx}& 0 & 0 \\
  0 & \ve_{yy} & 0 \\
  0 & 0 & \ve_{zz} \\
\end{array}%
\right),
\;\;\;\;\;\;\;\;\;\;
\hat{\mu}=\left(%
\begin{array}{ccc}
   \mu_{xx}& 0 & 0 \\
  0 & \mu_{yy} & 0 \\
  0 & 0 & \mu_{zz} \\
\end{array}%
\right).
$$
In the consider case the wave vector $\vec{k}$ is directed along the optical axis of the ULM $O$ and therefore the vectors of the electric $\vec{E}$  and  magnetic $\vec{H}$ fields and wave vector constitute a  left-handed triplet of vectors \cite{8}. For uniaxial anisotropy, the media are isotropic in the plane perpendicular to the optical axis $O$ and consequently  $\ve_{\perp}=\ve_{xx}=\ve_{yy}$ and $\mu_{\perp}=\mu_{xx}=\mu_{yy}$ and also take place the conditions $\ve_{||}\neq\ve_{\perp}$ and $\mu_{||} \neq \mu_{\perp} $, where  $\ve_{||}=\ve_{zz}$ and $\mu_{||}=\mu_{zz}$. The quantities $\ve_{||}, \;\mu_{||}$ and $\ve_{\perp},\;\mu_{\perp}$ are the  permittivity and permeability in the directions parallel and perpendicular to the optical axis $O$, respectively. Two cases are possible: if $\ve_{\perp}>\ve_{\perp}$, the crystal is  a negative uniaxial crystal, but if  $\ve_{\perp}<\ve_{\perp}$ of a positive one. For father consideration will be more convenient for the components of the permittivity tensor using denotations $\ve_{xx}$ and $\ve_{zz}$.

In optical region of spectra $d<<\lambda$, where $\lambda$ is length of the surface optical wave. Therefore, SIT can be model as SW propagation along the interface ULM/IRM and infinite small thickness transition layer (monolayer) with SQD \cite{23,25}.

For a surface TM-mode the electric field $\vec{E}(E_{x},0,E_{z})$ lies in the $xz$ plane perpendicular to the boundary of the division between two connected media and the magnetic field $\vec{H}(0,H_{y},0)$ is directed along the axis $y$.

For investigation of the SW  waves we  will follow to approach which is convenient for consideration  of the surface as linear so nonlinear waves as well \cite{23}. The quantity
\begin{equation}
U_{1}(x,z,t)=\int U_{1}(\Om,Q)e^{\kappa_{1}(\Om, Q)x}e^{i(Qz-\Om t)} d\Om
dQ,$$$$\;\;\;\;\;\;\;for\;\;\;\;x<0,
$$$$
U_{2}(x,z,t)=\int U_{2}(\Om,Q)e^{-\kappa_{2}(\Om, Q)x}e^{i(Qz-\Om t)} d\Om
dQ,$$$$\;\;\;\;\;\;\;for\;\;\;\;x>0,
\end{equation}
are a Fourier-decomposition of the fields. The quantities
\begin{equation}
\kappa_{1}^{2}(\Om, Q)=Q^{2}- \ve_{1}(\Om)\mu_{1}(\Om) \frac{\Om^{2} }{c^{2}},
\;\;\;\;\;\;\;\;\;\;\;\;$$$$
\kappa^{2}_{2}(\Om,Q)= \frac{ \ve_{zz}(\Om)}{\ve_{xx}(\Om)}Q^{2} - \mu_{yy}(\Om)\ve_{zz}(\Om) \frac{\Om^{2}}{c^{2}},
\end{equation}
characterize the transverse structure of the SW and from the Maxwell equations in connected media are determined.  The functions $U_{1,2}$ stands for the components $(E_{x},E_{z},H_{y},D_{x},D_{z},B_{y})$ in both connected media, where $D_{x},D_{z}$ and $B_{y}$ are the components of the displacement vector and the vector magnetic induction, respectively. We assume translational invariance in the $y$-direction, so that all field quantities  do not depend from the coordinate $y$, i.e. $\frac{\pa }{\pa y}=0.$

Taking into account the surface current caused by presence of the SQDs, the boundary conditions for surface waves at $x=0$ read \cite{23,25,30}:
\begin{equation}
H_{2,y}-H_{1,y}= \frac{4\pi}{c} \frac{\pa p}{\pa
t},\;\;\;\;\;\;{E}_{1,z}={E}_{2,z}.
\end{equation}
Using the eqs.(2)-(4)we obtain the nonlinear wave equation for the
$E_{z}$ component of the strength of electrical field at $x=0$, in
the following form \cite{24,26}:
\begin{equation}
\int f(\Om,Q)E(\Om,Q)e^{i(Qz-\Om t)} d\Om dQ=-4\pi p(z,t),
\end{equation}
where
\begin{equation}
 f(\Om,Q)=\frac{\ve_{zz} }{ {\kappa}_{2}} +  \frac{\ve_{1}}{{\kappa}_{1}},
\end{equation}
$E_{1,z}(\Om,Q)=E_{2,z}(\Om,Q)=E(\Om,Q)$. This equation is valid for any dependence of the polarization of the SQD $p(z,t)$ on the strength of electrical field at $x=0$. In order to determine the dependence of the polarization $p(z,t)$ on the strength of
electrical field at $x=0$, we have to consider the structure of the energetic levels of the SQD and the details of the nonlinear interaction of the surface pulse with the SQD. We assume that the SQD can be described by the ground $|0>$, exciton $|2>$ and
biexciton $|3>$ states.

The Hamiltonian of the system \cite{21,22,25}
$$
H=H_0 +V,
$$
where
$$
H_{0}=\hbar\omega_{12}|2><2|+\hbar\omega_{13}|3><3|
$$
is the Hamiltonian of the single-exciton and biexciton states and
$V=-\vec P  \vec E$ is the Hamiltonian of the light-quantum dots
interaction, $\hbar$ is the Planck's constant, $\om_{ij}$ are
frequencies of excitations between energetic levels $i$ and $j$
$\;(i,j,=1,2,3)$.

In general, the detunings from the resonance $\omega_{13} -
\omega_{12} - \omega$ and $\omega_{12} - \omega$, which describe the
SQDs, are different. Under the assumption of off-resonant excitation
with a constant detuning
$\omega_{13}-\omega_{12}-\omega\approx\omega_{12}-\omega=\Delta$ the
polarization which is determined by interband transitions occurring
in the quantum dots between the three energetic levels
\begin{equation}
p=N \int g(\Delta )[{\mu}_{12}\rho_{21}(\Delta)+{\mu}_{23}
\rho_{32}(\Delta) ] d \Delta +c.c.,
\end{equation}
where $N$ is the uniform quantum dot density,
 ${\mu}_{12}=\vec{\mu}_{12} \vec e,\;\;\;{\mu}_{23}=\vec{\mu}_{23}
\vec e; \;\;\vec{\mu}_{12}$ and $ \vec{\mu}_{23}$ are the dipole
elements for the corresponding transitions, $\vec{e}$ is the
polarization unite vector along $\vec{E}$. We assume the dipole
moments to be parallel to each other and to be directed along axis
$z$,$\;{\mu}_{12}={\mu}_{23}$; $g(\Delta \omega)$ is the
inhomogeneous broadening lineshape function resulting from dots size
fluctuations. The quantities $\rho_{ij}$ are the matrix elements of
the density matrix $\rho$ are determined by the Liouville equation
$$i\hbar\frac{\partial {\rho}_{nm}}{\partial
t}=\sum_{l}(<n|H|l>{\rho}_{lm}-{\rho}_{nl}<l|H|m>),
$$
where $n,m,l,=1,2,3$.

\vskip+0.5cm

\centerline{III. Solution of equations}

The solution of this equation we can find following the way
presented in the work \cite{18,32}. We can simplify Eq. (5) using the method
of slowly changing profiles. For this purpose, we represent the
functions $E$ and $p$ in the form
\begin{equation}
 E=\sum_{l=\pm1}\hat{E}_{l} Z_l;\;\;\;\;p=\mu_{12} N \tilde{p}\; Z_{1}+ c.c.
\end{equation}
where $\hat{E}_{l}$ and $\tilde{p}$  are the slowly varying complex
amplitudes of the optical electric field and the polarization,
$Z_{l}= e^{il(kz -\omega t)}$. To guarantee that $E$ is a real
number, we set $\hat{E}_{l}=\hat{E}_{-l}^{\ast}=\hat{E}$. This
approximation is based on the consideration that the envelopes
$\hat{E}$ vary sufficiently slowly in space and time as compared
with the carrier wave parts-i.e.,
\begin{equation}
|\frac{\partial \hat{E}}{\partial t}|<<\omega
|\hat{E}|,\;\;\;|\frac{\partial \hat{E}}{\partial z }|<<k|\hat{E}|,
\end{equation}
and analogously expressions for the $\tilde{p}$ is called the slowly varying envelope approximation \cite{32}.

Substituting the equations (6) and (8) in the wave equation (5), and
to take into account explicit form of the envelope of the
polarization (7) which determined from the Liouville equation, after
divided the real and imaginary parts of the equation (5) we obtain
dispersion law for surface pulse propagating
\begin{equation}
 f(\om,k)=0
\end{equation}
and a nonlinear wave equation in the form:
\begin{equation}\label{}
 (\frac{d \hat{E}}{d\zeta})^2 =T^{-2} {\hat{E}}^2  -\frac{{\mu_{12}}^2}{2{\hbar}^2 }
   {\hat{E}}^4,
\end{equation}
where  the width of the pulse $T$ is determined by the equation
\begin{equation}\label{}
T^{-2}=\frac{4\pi N {\mu}^{2}_{12}}{ f'_{\Om}(\frac{v_g}{V}-1)
\hbar}\int g(\Delta')F (\Delta')d \Delta'+O({\Delta}^2),
\end{equation}
$c$ is the speed of light in vacuum, $\zeta=t-\frac{z}{V}\;$, $V$ is
the constant pulse velocity,
\begin{equation}
v_{g}=-\frac{f'_{Q}}{f'_{\Om}},
\end{equation}
\begin{equation}
f'_{Q}=\frac{\pa f}{\pa Q}|_{\Om=\om,Q=k}=-Q(
\frac{\ve^{2}_{zz}}{\ve_{xx}\kappa^{3}_{2} }+
\frac{\ve_{1}}{\kappa^{3}_{1}})|_{\Om=\om,Q=k},
$$$$
f'_{\Om}=\frac{\pa f}{\pa \Om}|_{\Om=\om,Q=k}=$$$$=\{ \frac{\frac{d
\ve_{1}}{d \Om}}{ \kappa_{1}} + \frac{\ve_{1}\Om}{2c^{2}}\;\;
\frac{2 \ve_{1}\mu_{1}+\Om \frac{d (\ve_{1}\mu_{1})}{d \Om} }{
\kappa^{3}_{1}}+ \frac{\frac{d \ve_{zz}}{d \Om}}{ \kappa_{2}}- $$$$-
\frac{\ve_{zz}}{2}\;\; \frac{ \frac{d( \frac{\ve_{zz}}{\ve_{xx}})}{d
\Om}Q^{2} -\frac{\Om}{c^2}[2\mu_{\perp} \ve_{zz} +\Om
\frac{d(\ve_{zz}\mu_{\perp})}{d \Om} ]}{
\kappa^{3}_{2}}\}|_{\Om=\om,Q=k},
\end{equation}
where $\om$ and $k$ are frequency and wave number of the carrier wave, $
v_{g}$ is group velocity of linear SW.

The solution of Eq.(11) for the envelope function has the form \cite{18,26}
\begin{equation}
\hat {E}=\frac{2}{\mu_{0}T} sech{  \frac{t-\frac{z}{V}}{T}},
\end{equation}
where $\mu_{0}=\sqrt{2}\frac{{\mu_{12}}}{{\hbar} }$ and we can
determine the constant velocity of the $2\pi$ pulse of the SW:
\begin{equation}\label{}
\frac{1}{V}=\frac{1}{v_g}+\frac{4\pi N {\mu}^{2}_{12}}{\hbar
f'_{\Om} v_g }\int \frac{ g(\Delta') d
\Delta'}{T^{-2}+{{\Delta}'}^{2}}.
\end{equation}

In optical region of frequencies because the magnetic permeability
loses its usual physical meaning, in RM we must put magnetic
permeability $\mu_{1}=1$, otherwise would be an over-refinement\cite{30}.
We have to note that the description of an isotropic NIM in terms
$\ve(\om)$ and $\mu(\om)$ is not unique. Besides the $\ve(\om)$ and
$\mu(\om)$ approach for NIM, there is also an alternative
description,  based on the generalized, spatial dispersive
permittivity $\tilde{\ve}(\om,\vec{k})$. In this approach, the
non-local function $\tilde{\ve}(\om,\vec{k})$, depending besides
$\om$ also on the wave vector $\vec{k}$, describes both electrical
and magnetic responses of the medium and different optical effects
in NIM \cite{33}.

The Eqs. (2),(3),(10),(12)-(16) determine the parameters of the surface
soliton for any value of $x,z$ and $t$ and show that for the
existence of a soliton it is necessary that conditions $f'_{\Om}>0$
and $V < v_g$ are fulfilled. Parameters of the surface optical
solitons depend not only on the SQD parameters and the permittivities
of the two interface media, but  also depends on the magnetic
permeability $\mu_{\perp}(\om)$ and its derivative
$\frac{d\mu_{\perp}(\Om)}{d\Om}|_{\Om=\om}$ of the ULM.

At absence of a transition layer the nonlinear polarization $p=0$
and right hand side of the equation (5) equal to zero. In this linear limit the equation (5)
has simple solution with constant amplitude $U_{0}$. The parameters of
the linear SW propagating on the interface between ULM and IRM for any value of
$x,z$ and $t$, for the value of the a Fourier-factor
$U_{1,2}(\Om,Q)={U_{0;}}_{1,2}\delta{(\om-\Om)}\delta{(k-Q)}$, from the eqs.(2), (3), (10) and (13) are determined.

\vskip+0.5cm

\centerline{IV. Condition of existence of the SW}

The dispersion relation (10) for surface waves  propagating on the interface of the ULM and IRM interface are valid as for linear  so for nonlinear waves as well. Substituting the equations (3) in equation (10) the dispersion low for surface waves reduced to the form
\begin{equation}
 k^{2}=\frac{{\om}^{2}}{{c^2}}\frac{\ve_{xx}\ve_{1}(\ve_{zz}\mu_{1}-\ve_{1}\mu_{\perp}
)}{ \ve_{zz}\ve_{xx} -\ve^{2}_{1} }.
\end{equation}
At the determination of the dispersion law (10), have arose the term proportional of the integral $\int \frac{\Delta' g(\Delta') d\Delta'}{T^{-2}+{{\Delta}'}^{2}}$. In some special cases, under the condition when spatial dispersion is effective, this term is not small and it is influence on the dispersion law of the soliton of SIT significantly \cite{34}. But in the consider case this term is very small and in the theory of SIT always is neglected \cite{18,25,32}. Under this condition the dispersion low (17) for linear SW and surface soliton of SIT is the same and to take into account that $\mu_{1}=1$, reads:
\begin{equation}
 k^{2}=\frac{{\om}^{2}}{{c^2}}C(\om),
\end{equation}
where
$$
C(\om)=\ve_{xx}\ve_{1}\frac{\ve_{zz}-\ve_{1}\mu_{\perp}
}{ \ve_{zz}\ve_{xx} -\ve^{2}_{1} }.
$$
The surface waves can be exist only in the region of frequency $\om$
for which the quantities $\kappa_{1}$, $\kappa_{2}$ and $C$ are real and positive.

For the determination of the conditions of the existance of SW propagating on the interface ULH/IRH, we consider that for any value of the $\om$ for conventional media $\ve_{1}(\om)>0$ and for ULM $\mu_{\perp}(\om)<0$. From the dispersion relation (10) and equations (3) follows that for any value of $\om$ the both nonzero components of the electric permittivities of the ULM $\ve_{zz}(\om)<0$ and $\ve_{xx}(\om)<0$ are simultaneously negative.

Depending on the numerical values of the components of electric permittivity tensor of the ULM $\ve_{zz}(\om)$ and $\ve_{xx}(\om)$ two different conditions of the existence of the SW can take place:

The first condition:
\begin{equation}
\ve_{zz}<\ve_{1}\mu_{\perp},\;\;\;\;\;\;\;\;\; \ve_{zz}\ve_{xx}>\ve^{2}_{1}
\end{equation}
and the second one
\begin{equation}\ve_{zz}>\ve_{1}\mu_{\perp},\;\;\;\;\;\;\;\;\; \ve_{zz}\ve_{xx}<\ve^{2}_{1}.
\end{equation}

From the eqs.(19) and (20) we can see that the conditions of the existence of the surface waves in ULM/IRM  system depends not only from the signs of components of the electric permittivities and magnetic permeability ("rightness") but also on their numerical values and therefore in this system SW do not exists for any value of the frequency $\om$.

\vskip+0.5cm

\centerline {V. Total energy flow}

The time-averaged over a period $2\pi/\om$ of oscillation of the
field Poynting vector of the TM mode  $\langle \vec{S} \rangle$,
which is associated with the energy flow of the pulse, has the $z$
components in the IRM-medium 1 and ULM- medium 2 in the following form
\begin{equation}
\langle S_{1,z}\rangle =\frac{c^{2}k}{8\pi\om
\ve_{1}}|\hat{H}_{1}|^{2}e^{2
\kappa_{1}x},\;\;
$$$$
\langle
S_{2,z}\rangle =\frac{c^{2}k}{8\pi\om \ve_{xx}}|\hat{H}_{2}|^{2}e^{-2
\kappa_{2}x}.
\end{equation}
The corresponding total energy flow
\begin{equation}
N=\int_{-\infty}^{0}\langle {S_{1,z}}\rangle
dx+\int_{0}^{+\infty}\langle {S_{2,z}}\rangle dx=
$$$$
=\frac{\om C}{16\pi k}(\frac{|\hat{H}_{1}|^{2}}{\ve_{1}\kappa_{1}}+
\frac{|\hat{H}_{2}|^{2}}{\ve_{xx}\kappa_{2}}).
\end{equation}
To take into account that in ULM $\ve_{xx}=-|\ve_{xx}|$ the quantities
$\langle S_{1,z} \rangle$ and $\langle S_{2,z} \rangle$ have
opposite signs and consequently the surface TM-mode resonance
soliton will be have a vortexlike distribution of the Poynting
vector [35,36].

The equation (22) has the general form and valid for any boundary
condition for the strength of magnetic field of the SW. The total
energy flow is positive and coincide with the wave vector when the
condition is satisfied:
\begin{equation}
\kappa_{1}  \ve_{1}|\hat{H}_{2}|^{2}<\kappa_{2} |\ve_{xx}|
|\hat{H}_{1}|^{2}.
\end{equation}
In case when the transition layer is absent, the boundary condition
for envelopes of the strength of magnetic field of the surface
TM-mode has the following form: $ \hat{H}_{2}=\hat{H}_{1} $, and
because for SW $\ve_{1}<|\ve_{xx}|$ and $\kappa_{1}<\kappa_{2}$, the
condition (23) always is satisfied and total energy flow $N$ is
always positive [35,36].

Situation is changed when the resonance transition layer is
presented. From the boundary condition (4) the connection between $
\hat{H}_{2}$ and $\hat{H}_{1} $ has following form
\begin{equation}
|\hat{H}_{2}|^{2}= |\hat{H}_{1}|^{2}  + i R(\tilde{p}^{*}
\hat{H}_{1}- \tilde{p} \hat{H}^{*}_{1}) +R^{2}| \tilde{p} |^{2}
\end{equation}
where
$$
R=\frac{4\pi \om \mu_{12} N}{c}.
$$
From the equation (24) evident that the resonance transition layer
cause the change the total energy flow of the surface TM-mode and it
will be positive or negative depending from the parameter $R$ and
polarization $\tilde{p}$. When the condition
\begin{equation}
\kappa_{1} \ve_{1} [|\hat{H}_{1}|^{2}  + i R(\tilde{p}^{*}
\hat{H}_{1}- \tilde{p} \hat{H}^{*}_{1}) +R^{2}| \tilde{p}
|^{2}]<\kappa_{2} |\ve_{xx}|   |\hat{H}_{1}|^{2}
\end{equation}
is fulfilled the total energy flow will be positive (forward SW),
otherwise negative (backward SW).

In the considered case, the the quantities $ \hat{H}_{2}$ and
$\hat{H}_{1} $ are real functions and hence the equations (25) is
simplified
\begin{equation}
\kappa_{1} \ve_{1} [\hat{H}^{2}_{1}  + i R(\tilde{p}^{*} - \tilde{p}
)\hat{H}_{1} +R^{2}| \tilde{p} |^{2}]< \kappa_{2} |\ve_{xx}|
\hat{H}^{2}_{1}.
\end{equation}

Consequently, when the resonance transition layer is presented,
depending from the parameters of the SW, SQD and connected media,
the total energy flow can be have as positive so negative sign as
well.

\vskip+0.5cm

\centerline {VI.Numerical estimations}

In the optical NIM precious metallic nanostructures are used,
therefore the losses is significant for the surface optical waves in
LM. But in present work the questions regarding of the losses of
the surface optical modes and their amplifications are not
considered (see, for example \cite{37,38}). In metals and LM the SW
occurs when the carrier frequency $\om$ is below of plasma frequency
$\om_{p}$. For the coherent interaction of the pulse of surface wave
with medium the duration of the pulse  should be much shorter than
the characteristic plasmonic oscillation damping time $2\pi/ \gm$.
Silver is known as significantly lower losses than other metals at
the optical frequencies. Plasma frequency for silver $\om_{p}=2\pi
\times 2,18 \cdot 10^{15}\;s^{-1}$and damping frequency $\gamma=2\pi
\times 5,08 \cdot 10^{12}\;s^{-1}$ \cite{3}. On the other hand a planar
array of split-ring resonators can be fabricated upon gallium
arsenide (GaAs) substrate \cite{39}. On the surface or at the boundary of
semiconductors (for instant GaAs or InAs) with another medium, a
small concentration of SQD can be grown. Soliton of the SW in
regular media when one of the connected medium was semiconductor
with SQD is investigated \cite{25}. SQD is one of the promising object
also for amplification of the optical waves \cite{38,40}. Transverse
relaxation times of the quantum dots, which are of order of
nanoseconds to several tens of picoseconds \cite{41,42}, is longer than
the characteristic plasmonic oscillation damping time  $2\pi/ \gm$.
The envelope approximation (9) is appropriate for pulses when the
spectral width of the pulse much smaller than the $\om$ and valid
for pulses with width $T\gtrsim 20 fs$ \cite{43}. Consequently pulse with
width which is acceptable for investigation of  nonresonance nonlinear waves in the LHM \cite{15,16,17}, also satisfies  eq.(1) the conditions of SIT  in SQD . This circumstances allow to hope
that SIT in LM/RM with SQD transition layer can be experimentally
observable.

\vskip+0.5cm

\centerline {VII.Conclusion}

We considered the processes of propagation of the linear and nonlinear surface optical waves on the interface of the ULM and IRM.
In this system the linear and nonlinear surface waves have as common so different properties as well.

The common properties:

1. The dispersion low for surface waves determined by the equation(18).

2. Although in ULM some components of the electric permittivity and the magnetic permeability tensors can be as negative, so positive as well \cite{8}, it is shown that the surface waves can be formed only for the frequencies of the wave $\om$ for which the components of the electric permittivity and the magnetic permeability tensors of the ULM $\ve_{xx}, \;\ve_{zz},\;\mu_{\perp}$ are simultaneously negative but for IRM the permittivity $\ve_{1}$ is positive and permeability $\mu_{1}=1$. At the same time one of the conditions (19) or (20) is fulfilled.

3. The surface TM-modes have a vortexlike distribution of the Poynting vector.

4. The parameters of the surface waves depend not only on the permittivities of the two connected media, but (unlike the case of interface of the two conventional connected media \cite{25} ) also depends on the magnetic permeability $\mu_{\perp}(\om)$ and its derivative $\frac{d\mu_{\perp}(\Om)}{d\Om}|_{\Om=\om}$ of the ULM.

The surface linear wave and surface soliton have also different properies:

At the absent of the transition layer with SQD or when a carrier wave frequency $\om$ is far away  from the frequencies of excitations of the SQD, in this case SQD do not influence of the wave processes and under this condition linear SW is formed. The parameters of the linear surface waves for any value of $x,z$ and $t$, for the value of the a Fourier-factor
$U_{1,2}(\Om,Q)={U_{0;}}_{1,2}\delta{(\om-\Om)}\delta{(k-Q)}$, from the equations (2), (3), (10) and (13) are determined. The total energy flow $N$ is always positive.

Here we do not consider the case of the  linear resonance interaction of the SW with SQD which leads to the damping of the linear SW.

When a transition layer with SQD  sandwiched between the ULM and IRM the surface resonance soliton of SIT is formed. Explicit shape and parameters of the surface soliton (2$\pi$ pulse) for any value of $x,z$ and $t$  from the  Eqs.(2), (3), (10), (12)-(16) are determined. For the existence of a soliton it is necessary that conditions $f'_{\Om}>0$ and $V < v_g$ are fulfilled. Parameters of the surface optical solitons depend not only on the permittivities and  the magnetic permeability $\mu_{\perp}(\om)$ and its derivative  $\frac{d\mu_{\perp}(\Om)}{d\Om}|_{\Om=\om}$ of the LM but also depends from the SQD parameters. Depending from the parameters of the SW, SQD and connected media, the total energy flow can be have as positive (forward SW) so negative (backward SW) sign as well depending from the parameter $R$ and polarization $\tilde{p}$.

We have studied the $2\pi$ pulse but under the condition of SIT can be formed also a zero-area pulse (breather), when $|\Theta|<<1$ \cite{21,22}.  Besides with these pulses, a somewhat similar steady-state pulse propagation may be formed also in an amplifier medium with losses and $\pi$-pulse can be formed. In addition, in some special situations the SIT soliton laser is realized \cite{44}.

We consider case when transition layer contains SQD, but presented
results has more general characters and in particular case are valid also  when
transition layer contains atomic systems sandwiched between ULM/IRM.

We consider the case when optical axis $O$ is directed along the axis $z$ but all results are valid for the case when optical axis is directed along axis $x$ perpendicular of the interface, i.e. when $\ve_{xx}=\ve_{||}$ and $\ve_{zz}=\ve_{\perp}$.

In particular case, for the SW propagating on the interface of  isotropic LM and IRM
the equations (21)-(26) are also valid if in these equations for the ULM we make the following changes: $\ve_{zz}=\ve_{xx}=\ve_{2}<0,\;\;\mu_{\perp}=\mu_{2}<0$.
At this when $|\ve_{2}|>\ve_{1}$ the  surface waves exists when
$\ve_{2}<\ve_{1}\mu_{2}$, and in case when $|\ve_{2}|<\ve_{1}$ the  surface waves exists when
$\ve_{2}>\ve_{1}\mu_{2}$.

The investigated here the resonance nonlinear SW in LM, along with nonresonance nonlinear SW which have been considered before in LM \cite{15,16,17}, gives more complete representation about physical properties of nonlinear SW in LM.

Acknowledgements

A.G.T.would like to thank  the Max-Planck-Institut  f\"ur Physik Komplexer
Systeme and Technical University of Berlin for  hospitality, ISTC for financial support.

\end{document}